\documentclass[pra,aps,amsmath,amssymb,amsfonts,twocolumn,nofootinbib,floatfix]{revtex4}
\usepackage{}
\usepackage{amssymb}
\usepackage{bm,mathrsfs}
\usepackage{graphicx}
\usepackage{epsfig}
\usepackage{amsmath,bbm}
\usepackage{amsfonts,amssymb}
\usepackage{times}
\usepackage{verbatim}
\usepackage[sort&compress]{natbib}
\usepackage{amsmath}
\usepackage{bm}
\usepackage{float}
\usepackage{textgreek}
\usepackage{textcomp}
\allowdisplaybreaks[4]
\usepackage[colorlinks,breaklinks,linkcolor=blue,anchorcolor=blue,citecolor=blue,urlcolor=magenta]{hyperref}

\usepackage{color}
\definecolor{zzz}{rgb}{0.9,0.0,0.4}

\begin{document}

\title{Bound-State Engineered Quantum Batteries Against Decoherence}

\author{J. L. Li,$^{1,2}$ C. Cui,$^{1}$ W. Y. Hu,$^{1}$  Cheng Shang,$^{3,}$\footnote{\textcolor{zzz}{Corresponding author: cheng.shang@riken.jp}} Yan-Hui Zhou,$^{4,}$\footnote{\textcolor{zzz}{Corresponding author: yanhuizhou@126.com}} and H. Z. Shen$^{1,}$\footnote{\textcolor{zzz}{Corresponding author: shenhz458@nenu.edu.cn }}}
\affiliation{$^1$Center for Quantum Sciences and School of Physics, Northeast Normal University, Changchun 130024, China\\
$^2$Research Center for Quantum Physics and Technologies, Inner Mongolia University, Hohhot 010021, China\\
School of Physical Science and Technology, Inner Mongolia University, Hohhot 010021, China\\
$^3$Analytical Quantum Complexity RIKEN Hakubi Research Team,
RIKEN Center for Quantum Computing (RQC), 2-1 Hirosawa, Wako, Saitama 351-0198, Japan\\
$^4$Quantum Information Research Center and Jiangxi Province Key Laboratory of Applied Optical Technology, Shangrao Normal University, Shangrao 334001, China}
\date{\today}

\begin{abstract}
Quantum batteries promise revolutionary advantages for energy storage but are fundamentally crippled by environmental decoherence, which induces self-discharge and rapid ``aging." Here, we crack this critical bottleneck by exploiting the decoherence-suppression mechanism: the formation of system-environment bound states. We consider a charger-battery system embedded in a three-dimensional anisotropic photonic crystal bath and derive exact non-Markovian dynamics. Strikingly, we reveal a tunable phase diagram where modulating the atomic eigenfrequency and coupling strength switches the system between zero, one, and two bound states. In the presence of two bound states, the battery energy evolves into a persistent periodic oscillation, enabling lossless energy storage and on-demand extraction indefinitely—effectively realizing an aging-free quantum battery. Conversely, the absence of bound states leads to complete energy decay. We also demonstrate that even during self-discharge, a single bound state can stabilize extractable energy. This work establishes bound-state formation as a powerful and feasible strategy for combating decoherence, offering a clear blueprint for designing durable solid-state quantum batteries in non-Markovian photonic platforms.
\end{abstract}

\maketitle
\section{Introduction}
The development in the domain of quantum information \cite{Jaeger2007,Nielse2010} has raised great expectations. Quantum effects such as entanglement can be used to perform certain tasks and have significant advantages over classical devices \cite{Campaioli031001}, opening up new prospects for quantum computing \cite{Nielse2010,Rodriguez042618}, communication \cite{Gisin165}, measurement \cite{Li042629}, and other quantum technologies \cite{Riedel2030501,Acin20080201,Chen5,Chen033603,Zhou053718,Zhou2400089,Zhou064009,Sun043715,Xue4424,Yang023703,Luan2350021,Shen2350029,Wang64003,Shen023849,zz4}. Quantum batteries (QBs) \cite{Song020405,Andolina240403,Medina220402,Gyhm128140501,Quach83160,Ukhtary123034001,Downing6322,Gemme843,Yang5102300,Gyhm6012001,Shaghaghi25430,Santos107032203,Kamin56275302,Gumberidze919628,Crescente22063057,Joshi042601,Yu062614,Lu043706,Yao044116,Song054107,Song090401,Zhang054125,
Wang062402,Huang030201,Xu012425,Friis261,Santos100032107,Barra122210601,Alicki150214110,Chen5321900487,Li5614,Rodriguez042419,
Mitra012227,Centrone052213,Sen030402,
Downing052206,Bhanja012224,Gemme023091,Bakhshinezhad014131,Konar042207,Almeida052218,Mazzoncini032218,Imai022215,Santos052203,
Ghosh022628,Qi032606,Gherardini191002458,Quach14024092,Santos101062114,Kamin102052109,Zozulya88043641,Wurtz95056802,Bai060201} as a research on technology miniaturization of nanodevices \cite{Binder17075015,Campaioli118150601,Andolina98205423,Ahmadi210402,Caravelli2023095} were first proposed by Alicki and Fannes in 2013 \cite{Alicki87042123}. 
Despite their superiority over classical batteries \cite{Vincent1997,Dell2001,Patil2043196} in terms of efficient charging rates \cite{JuliaFarre181104005,Ghosh032207,Lai023136,Seah100601,Shrimali022425}, superior power extraction \cite{Hovhannisyan111240401,Song064103,Hao012207,Tirone060402,Francica044119,Hoang013038,Bhattacharyya240401,Andolina122047702,Giorgi48035501,GarciaPintos190903558,Francica312,
Fusco94052122,Monsel124130601,Struchtrup120250602}, and utilization of collective quantum  properties \cite{Zhang032211,Yang062432,Gao043150,Ferraro120117702,Andolina99205437,Rossini100115142,Tomadin27130,Carrasco064119,Francas032205,Kamin022226,Mondal044125,Shukla260303883,Shukla260103119} to store energy, significant efforts remain focused on enhancing their performance \cite{Ma022433,Zhao013172,Yang030402,Arjmandi062609,Zhang042424,Yang012204} and overcoming inevitable challenges. Since then, a variety of possible QBs have been constructed, including Dicke batteries \cite{Seidov022210,Crescente245407}, Spin-chain and spin-network batteries \cite{Mojaveri042619,Salvia013155,Catalano030319,Zhao033715,Le97022106,Wang69205312,Xie10034005,Yu8054038,Mahfouzi98205423,Yao014138,Chen054119,Peng052220,
Liu042411,Liu245418},  Sachdev-Ye-Kitaev (SYK) quantum batteries \cite{Rossini125236402}, ordered and disordered quantum batteries \cite{Arjmandi064106,Ghosh101032115}, etc.

Particularly, a QB model that is minimal but popular and based on a two-level system has been intensively explored in both theoretical research and experimental realization \cite{FornDiaz39,Yan180401}. A related challenge is the decoherence problem of quantum batteries, which affects charging process, energy storage and extraction \cite{Morrone044073,Pirmoradian100043833,Zakavati200309814,Carrega22083085,Xu064143,Lu03675}. By coupling with a charger, a battery can be isolated from the environment and regarded as a non-dissipative subsystem to prevent energy loss \cite{Andolina98205423,Farina99035421}. The charger supplies energy to the battery in what is called the charging process. Once disconnected, the battery can store energy, which can then be extracted for utilization \cite{Farina99035421,Abah022017}. Recent research findings show that when a qubit (QB) and a quantum charger are connected to a reservoir, like a rectangular hollow metal waveguide, it enables effective remote charging of the qubit. However, this connection unavoidably leads to a low level of stored energy and a reduction in ergotropy \cite{Song19226}. The decoherence problem caused by the environment still affects the performance of the battery and triggers the self-discharge process, i.e., aging of QB \cite{Kamin22083007,Zhu391978998}.

In the field of quantum information, many suppression schemes have been proposed for the decoherence problem of nanoscale solid-state devices, such as those induced by non-Markovian coherent feedback control \cite{Xue052304,Zhang6791}, dynamic decoupling control \cite{Viola2733}, and bound state formation \cite{John2418,Shen93012107,Shen99032101,Shen98062106}, where the efforts to suppress decoherence in non-Markovian systems mainly depend on the physics of a single two-level system in a structured environment. The prediction that the system-environment bound state can restrain decoherence caused by environmental feedback has been verified in the recent circuit Quantum Electrodynamics (QED) experiment \cite{Liu48,Krinner559,Sundaresan011021}. Inspired by this, we propose coupling the charger and battery to a 3D anisotropic photonic crystal, creating a non-Markovian structured environment to suppress decoherence for the quantum battery subjected to the external environment.

Despite these advances, a central missing link remains: a controllable strategy to prevent long-term energy loss beyond the Markovian limit. Most existing schemes either rely on dynamical decoupling, which requires active external driving, or focus solely on the short-time charging power, ignoring the inevitable long-time aging. In this work, we provide exactly the missing link. We demonstrate that passively engineering the reservoir spectrum to host bound states offers an unprecedented and intrinsic handle to tame dissipation. Moving beyond the single-atom paradigm, we couple a two-qubit charger-battery system to a 3D anisotropic photonic crystal. By analytically solving the integro-differential equations via complex contour integration, we uncover a rigorous connection between the number of bound states (zero, one, or two) and the battery's dynamical phases. This yields a concrete, experimentally accessible phase diagram. We find that the two-bound-state regime fundamentally overcomes the aging problem, while the single-bound-state regime significantly enhances steady-state ergotropy.

The remainder of this paper is organized as follows. In Sec.~\ref{Model and Dynamics}, we introduce the model and derive a non-Markovian master equation for QB subjected to the coupling of the charger and 3D anisotropic photonic crystal environment. In Sec.~\ref{Quantum Battery in Non-Markovian Environment}, we first derive the short-term exact solution of system dynamics and demonstrate the influences of different bound state numbers on the dynamical behavior of the QB in a non-Markovian environment. The threshold of the tunable parameters for forming bound states is also revealed. Moreover, we investigate long-term charging and energy extraction and propose an optimal scheme. Sec.~\ref{Self-Discharging Process of Quantum Battery} studies the self-discharging process of QB, demonstrating the ability of the bound state to resist decoherence within the constructed non-Markovian environment. Finally, we conclude in Sec.~\ref{Conclusions and Discussions}.

\section{Model and Dynamics}\label{Model and Dynamics}

\begin{figure}[t]
\centerline{
\includegraphics[width=8.1cm, height=5.6cm, clip]{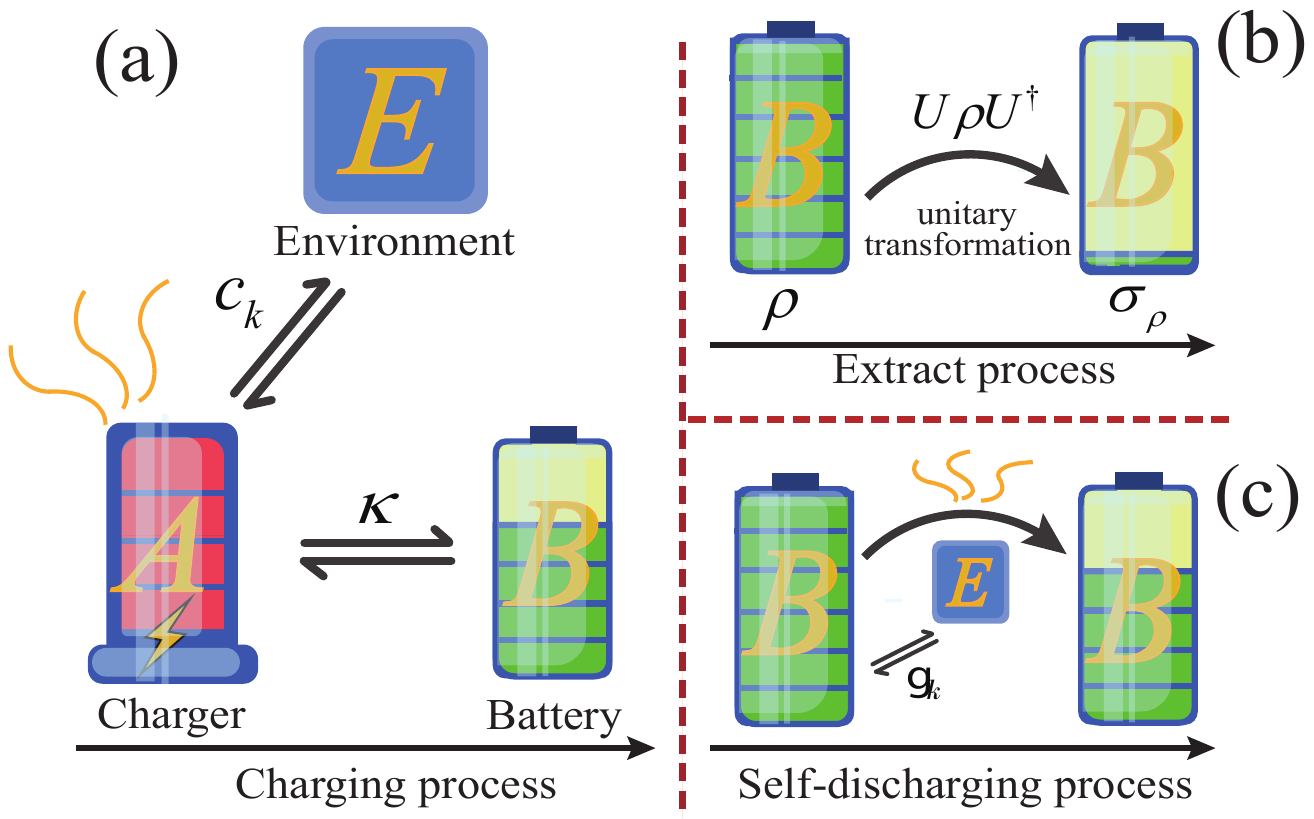}}
\caption{QB models against decoherence. Setup: (a) Charging process of two two-level atoms considered as charger $A$ and QB $B$ respectively in an amplitude damping environment $E$ composed of a 3D anisotropic photonic crystal. QB is isolated from the environment by a charger. Here, ${{c_k}}$ represents the coupling strength between the charger and the structured environment, while $\kappa $ denotes the interacting coefficient between the charger and the battery. (b) Unitary process to extract available energy in QB $B$. This means the battery is decoupled from the charger after fully charging. The battery state undergoes a unitary evolution process from $\rho $, i.e., $U\rho {U^\dag }$, resulting in the extraction of all available energy, where the final state is transformed into a passive state ${\sigma _\rho }$. (c) Self-discharge process of a fully charged QB in the same amplitude damping environment $E$ as in (a). When the battery is fully charged and stored instead of being extracted, a self-discharge process will occur, i.e., energy dissipation loss due to the decoherence effect caused by the environment.} \label{Installation}
\end{figure}
\begin{figure}[t]
\centerline{
\includegraphics[width=7.8cm, height=3.5cm, clip]{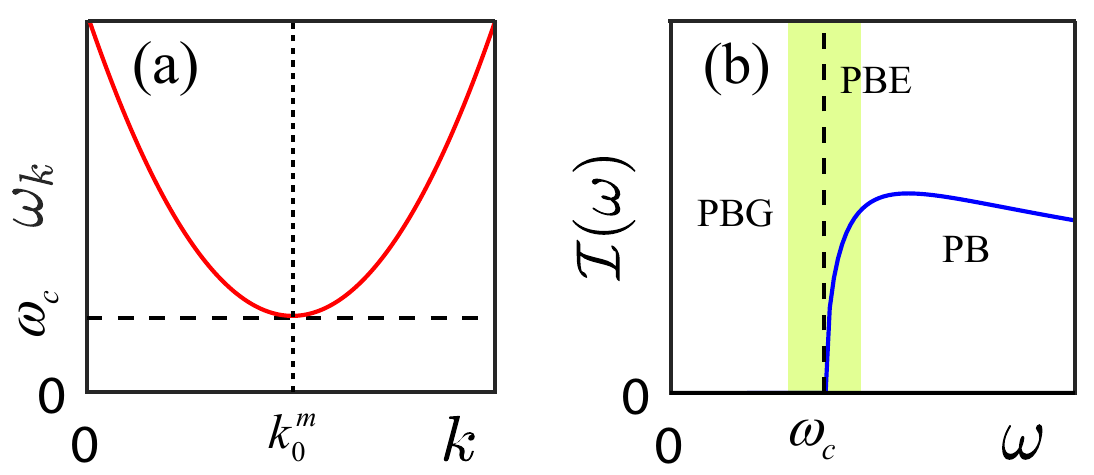}}
\caption{(a) The dispersion relation (\ref{wk}) with the photonic band edge ${\omega _c}$ for the system. (b) shows the band structures of the photonic crystal environment. The spectral density  ${\cal{I}}(\omega )$ in Eq.~(\ref{Jw}) for a two-level atom is coupled to a photonic crystal. The spectrum is classified into three regimes: photonic band gap (PBG), photonic band edge (PBE), and the photonic band (PB). } \label{PD}
\end{figure}

We study a system composed of two coupled two-level atoms (two qubits) with the atomic levels ${\left| e \right\rangle_A}$, ${\left| g \right\rangle_A}$, ${\left| e \right\rangle_B}$, and ${\left| g \right\rangle_B}$, where the eigenfrequency of each atom is ${\omega _0}$. One is considered as a QB $B$, the other is regarded as a charger $A$, which couples with a structured environment $E$ as shown in Fig.~\ref{Installation} (a). Here, the environment is the 3D anisotropic photonic crystal \cite{Bayindir842140} as depicted in Fig.~\ref{PD}. The Hamiltonian of the whole system can be expressed as (with $\hbar  = 1$) \cite{Bernier023841,Aspelmeyer1391,Mahmoodian133603}
\begin{equation}
\begin{aligned}
\hat H= {{\hat H}_0} + {{\hat H}_I}
,\label{Ht}
\end{aligned}
\end{equation}
with
\begin{align}
{\hat H_0} &= {\rm{ }}{\omega _0}\hat \sigma _A^ + \hat \sigma _A^ -  + {\omega _0}\hat \sigma _B^ + \hat \sigma _B^ -  + \sum\limits_k {{\omega _k}} \hat a_k^\dag {\hat a_k},\label{H0}\\
{{\hat H}_I} &= \kappa (\hat \sigma _A^ + \hat \sigma _B^ -  + \hat \sigma _A^ - \hat \sigma _B^ + ) + \sum\limits_k {({c_k}} \hat \sigma _A^ + {\hat a_k} + c_k^ * \hat \sigma _A^ - \hat a_k^\dag ),\label{H0HI}
\end{align}
where $\hat \sigma _A^ + $($\hat \sigma _B^ + $) and $\hat \sigma _A^ - $($\hat \sigma _B^ - $) are the Pauli raising and lowering operators for atom $A$($B$). Here, ${\hat a_k}$ ($\hat a_k^\dag $) denotes the annihilation (creation) operator of the $k$th field mode with frequency ${{\omega _k}}$, while $\kappa $ represents the coupling strength of the interaction between two atoms. The coupling strength between the atom $A$ and the $k$th field mode is described by ${{c_k}}$. Hamiltonian (\ref{Ht}) is analytically solvable due to the total excitation number $\hat N = \hat \sigma _A^ + \hat \sigma _A^ -  + \hat \sigma _B^ + \hat \sigma _B^ -  + \sum\nolimits_k {\hat a_k^\dag {{\hat a}_k}} $ being conserved. In Hamiltonian~(\ref{H0HI}), the first term represents the linear coupling between two-level systems, whereas the second term describes either (i) the interaction between a two-level system with a circularly polarized dipole moment and a co-polarized environmental mode~\cite{Mahmoodian133603}, or (ii) the attractive interaction between the system and the environmental modes~\cite{Bernier023841}. For simplicity, we calculate the one-excitation time evolution of the total system when the environment is initially in a vacuum state. Hence, the total system dynamics can be written as a superposition given by
\begin{equation}
\begin{aligned}
\left| {\psi (t)} \right\rangle & = [{{\cal R}_A}(t){\left| e \right\rangle _A}{\left| g \right\rangle _B} + {{\cal R}_B}(t){\left| g \right\rangle _A}{\left| e \right\rangle _B}] \otimes {\left| {{0_k}} \right\rangle _E}\\
&\quad+ \sum\limits_k {{{\cal G}_k}} (t){\left| g \right\rangle _A}{\left| g \right\rangle _B} \otimes {\left| {{1_k}} \right\rangle _E},
\label{psit}
\end{aligned}
\end{equation}
with probability amplitudes ${{\cal R}_A}(t)$, ${{\cal R}_B}(t)$, and ${{\cal G}_k}(t)$ being functions to be determined. While the probability amplitude ${{\cal R}_A}(t)$  corresponds to the total system to be in the combined state where subsystem $A$ is excited (${\left| e \right\rangle _A}$), subsystem $B$  is in the ground state (${\left| g \right\rangle _B}$), and mode $k$ of the field  $E$ has no photon (${\left| {{0_k}} \right\rangle _E}$). The other probability amplitudes and states have similar notations. According to the Schr\"{o}dinger equation, the probability amplitude for atom $A$ is given by the following integro-differential equation (see Appendix~\ref{A} for details)
\begin{equation}
\begin{aligned}
\frac{d}{{dt}}{{\cal R}_A}(t) &=  - i{\omega _0}{{\cal R}_A}(t) - \int_0^t {{\cal M}(t - \tau ){{\cal R}_A}(\tau )} d\tau\\
&\quad- \int_0^t  {\cal F}(t - \tau ){{\cal R}_A}(\tau )d\tau
,\label{DRAt}
\end{aligned}
\end{equation}
where ${\cal M}(t- \tau) = {\kappa ^2}{e^{ - i{\omega _0}(t - \tau )}}$. The correlation function is ${\cal F}(t - \tau ) = \int {\cal I} (\omega ){e^{ - i{\omega}(t - \tau )}}d\omega$ with the spectral density of the structured environment given by
\begin{equation}
\begin{aligned}
{\cal I}(\omega ) = \sum\limits_k {\left| {c_k} \right|^2} \delta (\omega  - {\omega _k})
,\label{jw}
\end{aligned}
\end{equation}
which characterizes all the interactions between atoms and the photonic crystal. Moreover, it can be uniquely determined by the coupling strength ${\left| {{c_k}} \right|^2}$ between atoms and the photonic crystal. The second term and third term of Eq.~(\ref{DRAt}) denote the influences of the battery and 3D anisotropic photonic crystal environment on the dynamics for the charger, respectively. One is that the battery drives the quantum state of the charger to evolve toward coherence, while the other is that the 3D anisotropic photonic crystal environment causes the charger to generate dissipation with non-Markovian photon backflow.

For the 3D anisotropic photonic crystal environment, the coupling coefficient ${c_k}$ takes \cite{John1764}
\begin{equation}
\begin{aligned}
{c_k} = ({\omega _0}d)\sqrt {1/(2{\varepsilon _0}{\omega _k}V)} \bm{e_k} \cdot \bm{u}
,\label{ck}
\end{aligned}
\end{equation}
where $k$ denotes both the momentum and the polarization of the modes. ${{\varepsilon _0}}$ is the vacuum dielectric constant. The magnitude and unit vector of the atomic dipole moment of the transition are denoted by $d$ and $\bm{u}$, respectively. $V$ stands for the quantization volume. $\bm{{e_k}}$ represents the transverse unit vector for the environment modes. In the case of a photonic-crystal-like environment, the dispersion relation reads \cite{John2486}
\begin{equation}
\begin{aligned}
{\omega _k} = {\omega _c} + {\cal A}{\left| \bm{k}  - \bm{k_0^{(m)}} \right|^2},
\label{wk}
\end{aligned}
\end{equation}
as shown in Fig.~(\ref{PD})(a), where ${\omega _c}$ is the cutoff frequency of the band edge, and $\bm{k_0^{(m)}}$ denote the finite collections of symmetry-related points that are associated with the band edge. The parameter ${\cal A}$ is a model-dependent constant. By utilizing the dispersion relation (\ref{wk}), converting the summation of modes over transverse plane waves into an integral, and then performing the integral operation, we analytically obtain the spectral density of the structured environment
\begin{equation}
\begin{aligned}
{\cal I}(\omega ) = \frac{{{\Gamma ^{3/2}}}}{\pi }\frac{{\sqrt {\omega  - {\omega _c}} }}{\omega }\Theta  (\omega  - {\omega _c}),
\label{Jw}
\end{aligned}
\end{equation}
as shown in Fig.~(\ref{PD})(b), where ${\Gamma ^{3/2}} = {({\omega _0}d)^2}\sum\nolimits_{{m}} {{{\sin }^2}({\theta _{{m}}})/(4\pi {\varepsilon _0}{{\cal A}^{3/2}})} $ with ${{\theta _{{m}}}}$ being the angle between the dipole vector of the atom and the $m$th $\bm{k_0^{(m)}}$. Here, $\Gamma $ represents the effective coupling strength between the atomic system (charger) and the structured environment, and $\Theta  (\omega  - {\omega _c})$ is an unit step function, i.e., $\Theta (\omega  - {\omega _c}) = 1$ for $\omega  \ge {\omega _c}$; otherwise, $\Theta (\omega  - {\omega _c}) = 0$.  In the next Sec.~\ref{3.1}, we will solve the Eq.~(\ref{DRAt}) to get the exact solution of ${{\cal R}_A}(t)$ in order to determine the QB energy and ergotropy. On the other hand, the dynamics of QB is governed by the exact non-Markovian master equation by tracing over the degrees of freedom of the charger and environment
\begin{equation}
\begin{aligned}
{{\dot \rho }_B}(t) = \gamma (t)[\hat \sigma _B^ - {\rho _B}(t)\hat \sigma _B^ +  - \frac{1}{2}\hat \sigma _B^ + \hat \sigma _B^ - {\rho _B}(t) - \frac{1}{2}{\rho _B}(t)\hat \sigma _B^ + \hat \sigma _B^ - ]
,\label{Drho}
\end{aligned}
\end{equation}
where the dissipation rate
\begin{equation}
\begin{aligned}
\gamma (t) =  - 2{\mathop{\rm Re}\nolimits} [{{\dot {\cal R}}_B}(t)/{{\cal R}_B}(t)]
.\label{gamma1}
\end{aligned}
\end{equation}
In the following section, we will show dynamics of QB and discuss the formation of bound states and influence for the whole system.

\section{QB in Non-Markovian Environment}\label{Quantum Battery in Non-Markovian Environment}
\begin{figure}[t]
\centerline{
\includegraphics[width=6cm, height=7.2cm, clip]{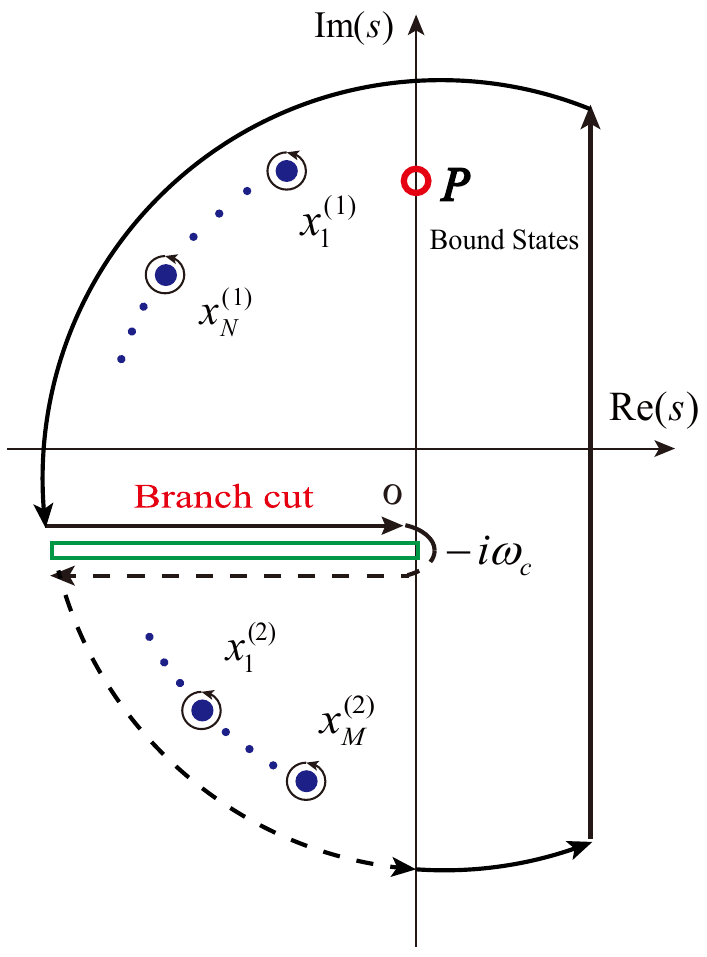}}
\caption{The anisotropic 3D photonic crystal environment revolves around the band edge at \(s=-i\omega_c\) to avoid branch cutting. The figure shows the situation in the complex plane \(s\), where the green line is the branch cut; the point \(P\) represents the bound state. The integration along the solid (dashed) curve is made on the first (second) Riemannian sheet. On the first Riemann sheet, the poles are labeled as \(x_{1}^{(1)},\cdots,x_{N}^{(1)}\); on the second Riemann sheet, the poles are labeled as \(x_{1}^{(2)},\cdots,x_{M}^{(2)}\) in the ${\mathop{\rm Re}\nolimits} (s) < 0$ half plane.} \label{s}
\end{figure}
In the case of a single atom coupled with the environment, the whole system (system plus its environment) can form a bound state, which is essentially a stationary state with a vanishing decay rate during its time evolution \cite{Liu48}. If such a bound state is established, it will result in dissipationless dynamics. In this section, it is demonstrated that the non-Markovian dynamics of an open system is closely related to the energy-spectrum characteristics of the whole system (including the charger-battery system and its environment). Consequently, the examination of the energy spectrum can offer insights for understanding its dynamics for the decoherence suppression.  We note that the system-environment bound states can be easily controlled by manipulating the coupling strength of the interaction between two atoms and the atomic eigenfrequency. Now we first present the exact solution of the non-Markovian dynamics for the system.\cite{Luan2503,Zhang1090337012024,Sun063713,Cui032209,Sun063704,tang1100437062024,Cui042129,Shen315042019,Shen28522018,Li109023712,shen880338352013,shen1050237072022,Shen1070537052023,Shen042129,shen950121562017,ShenNHgain2025,Xin105053706,Yang1090537122024,Wang34001,Shene71535,shen960338052017}
\begin{subsection}{The exact solutions of the system dynamics}\label{3.1}

Employing complex multivalued function contour integration and the residue theorem, we solve Eq.~(\ref{DRAt}) and obtain the exact expression of the probability amplitude for atom  $A$
\begin{equation}
\begin{aligned}
{R_A}(t) &= {R_A}(0) \Bigg\{  \sum_M \frac{e^{x_M^{(1)} t}}{P'(x_M^{(1)})} 
+ \sum_N \frac{e^{x_N^{(2)} t}}{Q'(x_N^{(2)})} \\
&\quad + \frac{1}{2\pi i} \int_0^\infty du \, h_1(-u - i\omega_c) e^{-ut - i\omega_c t} \Bigg\}
,\label{RAt1}
\end{aligned}
\end{equation}
 where ${P'(x_M^{(1)})}$ and ${Q'(x_N^{(2)})}$ are the results of substituting ${x_M^{(1)}}$ and ${x_N^{(2)}}$ into the derivative of $P(s)$ and $Q(s)$ with respect to $s$. Here, $P(s)$ and $Q(s)$ are given by Eq.~(\ref{Ps}) and Eq.~(\ref{Qs}), and $s$ is a complex variable that represents the frequency domain in the Laplace transform. Also, ${x_M^{(1)}}$ and ${x_N^{(2)}}$ are the roots of equation $P(s) = 0$ and $Q(s) = 0$, respectively. The function ${h_1}(s) = {Q^{ - 1}}(s) - {P^{ - 1}}(s)$ and the variable $u =  - s - i{\omega _c}$ are defined. The calculation details of Eq.~(\ref{RAt1}) can be found in Appendix \ref{D}.

The first term in Eq.~(\ref{RAt1}) represents localized modes with $s =  - iE$  ($E$ is real, corresponding to the eigenenergy spectrum of the whole system). Localized modes exist only if the environmental spectral density has band gaps at pure imaginary zeros where $P( - iE) = 0$ gaps located [see point $P$ in Fig.~(\ref{s})]. These localized modes don't decay, giving dissipationless non-Markovian dynamics. The nonlocalized mode has two parts: one is the second term in Eq.~(\ref{RAt1}), an oscillating damping process due to complex roots in $Q(s) = 0$  in the regime of [${\mathop{\rm Re}\nolimits} (s) < 0$ and ${\mathop{\rm Im}\nolimits} (s) <  - {\omega _c}$], while the other is the integral part where nonexponential parts oscillate rapidly in time, which comes from the term with ${e^{i{\omega _c}t}}$ in Eq.~(\ref{RAt1}). The nonlocalized mode part in equation is the contribution of allowed bands that usually generate exponential decays, which is another significance of non-Markovian dynamics. It is straightforward to show the battery probability amplitude ${{\cal{R}}_B}(t)$ according to the integral equation (\ref{RB}). Based on the reduced density operator ${\rho _B}(t)$ of the QB, the battery energy
\begin{equation}
\begin{aligned}
{E_B}(t) = {\omega _0}{\left| {{{\cal R}_B}(t)} \right|^2}
\label{EB}
\end{aligned}
\end{equation}
is obtained by ${E_B}(t) = {\rm{Tr[}}{\rho _B}(t)\hat H_0^B{\rm{]}}$ with $\hat H_0^B = {\omega _0} \hat \sigma _B^ + \hat \sigma _B^ - $.
\begin{figure}[t]
\centering
\includegraphics[width=0.48\textwidth, height=7.0cm, clip]{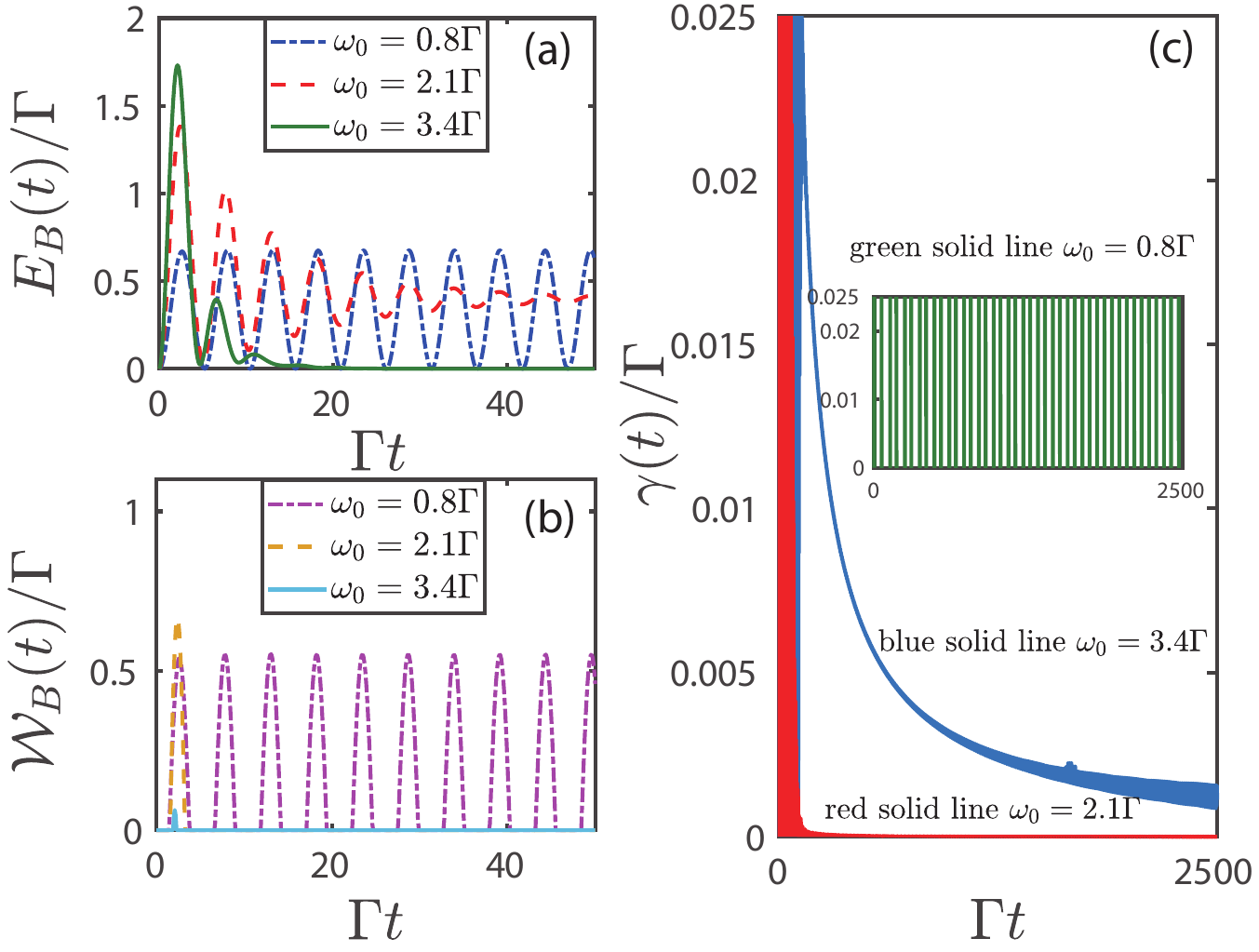}
\caption{(a) The time evolution of the battery energy ${E_B}(t)$ in Eq.~(\ref{EB}) as a function of the dimensionless quantity $\Gamma t$ with different ${\omega _0}$ (in units of $1/\Gamma $). Here, ${\omega _0} = 0.8\Gamma $ corresponds to the regime of forming two bound states. ${\omega _0} = 2.1\Gamma $ falls into the regime of forming one bound state. ${\omega _0} = 3.4\Gamma $ represents no bound state. (b) displays ${{\cal W}_B}(t)$ given by Eq.~(\ref{WB}), while (c) shows the corresponding decay rate $\gamma (t)$ provided by Eq.~(\ref{gamma1}) with and without bound state. All parameters are dimensionless, and the other chosen parameters are  $\kappa  = 0.59\Gamma $, ${\omega _c} = 2\Gamma $.}
\label{all}
\end{figure}
To characterize the maximal amount of energy that can be extracted from a QB at the end of the charging process under the cyclic unitary operations as shown in Fig.~\ref{Installation} (b), the ergotropy is introduced as \cite{Farina99035421}
\begin{equation}
\begin{aligned}
{{\cal W}(t)} = {\rm{Tr[}}{\rho(t)}{{\hat H}}{\rm{] - }}\mathop {{\rm{min}}}\limits_U {\rm{Tr[}}U{\rho(t)}{U^\dag }{{\hat H}}{\rm{]}}
,\label{WB1}
\end{aligned}
\end{equation}
in which ${\rho (t)}$ is the battery state when charging is completed at time $t$. The first term of Eq.~(\ref{WB1}) actually represents battery energy ${E_B}(t)$. In the second term, minimization is performed over all the formal unitaries $U$ acting locally on the battery, which means that all available energy of the battery is extracted. According to some definitions of ergotropy, we obtain
\begin{equation}
\begin{aligned}
{{\cal W}_B}(t) =  {\omega _0}[2{\left| {{{\cal R}_B}(t )} \right|^2} - 1]\Theta ({\left| {{{\cal R}_B}(t )} \right|^2} - \frac{1}{2})
,\label{WB}
\end{aligned}
\end{equation}
 whose derivation details can be found in the Appendix \ref{D}. Next, we will investigate the energy spectrum of the whole system to help us understand the decoherence suppression induced by the formation of bound states, which leads to the exact non-Markovian dynamics with dissipation inhibition for the charger and battery.

\end{subsection}

\begin{subsection}{Short-Term Behavior and Threshold in the Influences of Non-Markovian Environment on QB}

After obtaining the exact solutions of the system dynamics, we now turn to analyze the short-term behavior and threshold in the non-Markovian environment feedback on the QB. According to Eqs.~(\ref{EB}) and (\ref{WB}), the time evolution for battery energy and ergotropy can be obtained in Fig.~\ref{all}(a)(b). We plot three lines with ${\omega _0} = 0.8\Gamma $, ${\omega _0} = 2.1\Gamma $, and  ${\omega _0} = 3.4\Gamma $ in each figure and point out that ${\omega _0}$ at different regime will determine different dynamical behaviors. To be specific, when ${\omega _0} = 0.8\Gamma $, the dynamics attains periodic oscillation behaviors. It holds a non-zero steady value after a long period when ${\omega _0} = 2.1\Gamma $. However, when ${\omega _0} = 3.4\Gamma $, the battery energy is always rapidly approaching zero after a period of oscillation in Fig.~\ref{all}(a). For the ergotropy, according to Eq.~(\ref{WB}), we know that when ${\left| {{{\cal R}_B}(t)} \right|^2} > \frac{1}{2}$, the energy of the battery can be extracted, i.e., ${E_B}(t) > 0.5{\omega _0}$. Therefore, it can always extract energy when ${\omega _0} = 0.8\Gamma $ and can be seen that the energy cannot be extracted after a period of time when ${\omega _0} = 2.1\Gamma $ and  ${\omega _0} = 3.4\Gamma $. Concerning the quantity of extractable energy, the presence of bound states exhibits a distinct comparative advantage over the situation where bound states are absent.

The physics behind these phenomena is related to the formation of bound states and corresponding decay rate as shown in Fig.~\ref{all}(c) according to Eq.~(\ref{gamma1}). We plot the dissipation rate $\gamma (t)$ for different ${{\omega _0}}$. When ${\omega _0} = 0.8\Gamma $, the dissipation rate oscillates continuously, which leads to the constant oscillation of battery energy corresponding to the regime of forming two bound states. Time-dependent dissipation $\gamma (t)$ decays to zero with oscillations for ${\omega _0} = 2.1\Gamma $, which means that battery energy will oscillate and decay until it stabilizes, corresponding to the regime of forming one bound state. When \({\omega _0} = 3.4\Gamma\), the fact that the dissipation rate ultimately reaches a stable state implies that the energy of the battery will decline to zero. This situation is associated with the absence of bound states, while the formation of bound states is closely related to the energy-spectrum signatures of the whole system. Therefore, the investigation of the energy spectrum may provide us with insight to understand its dynamics.
\begin{figure*}[t]
\centering
\includegraphics[width=0.66\textwidth, height=7.8cm, clip]{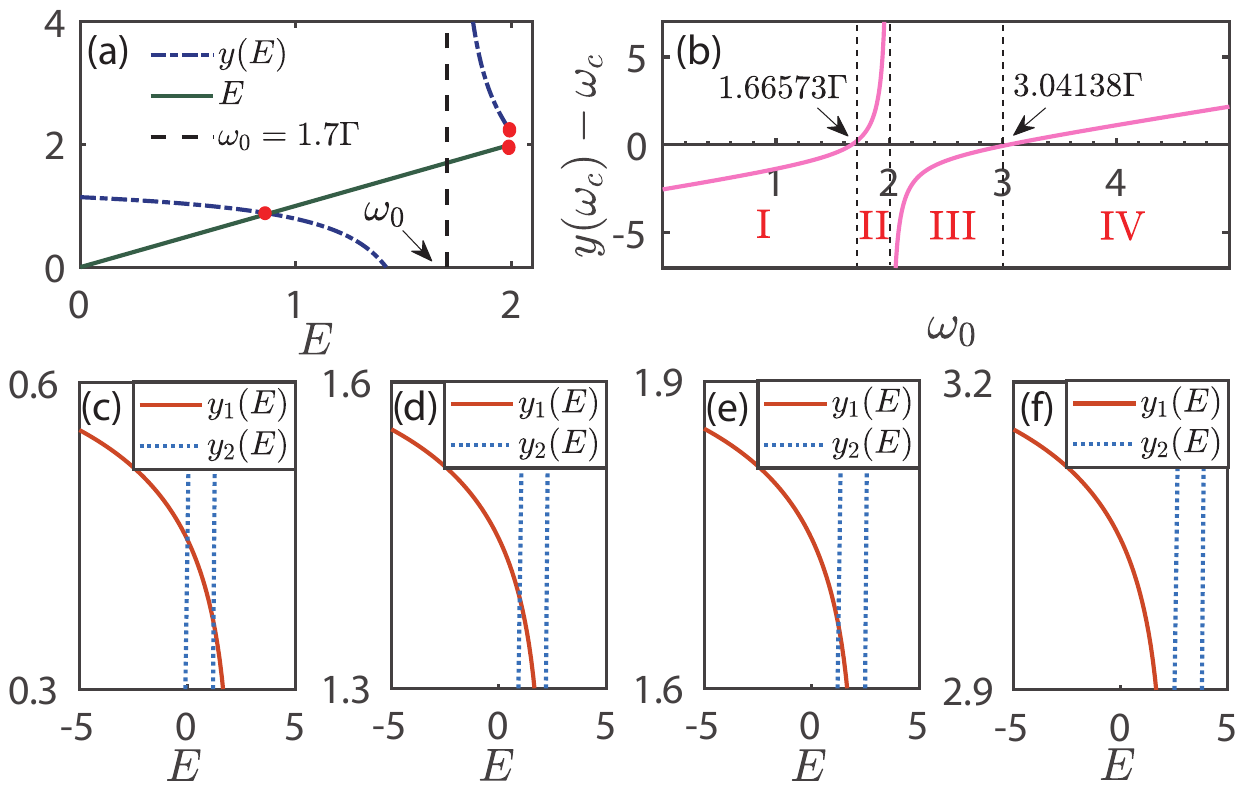}
\caption{(a) shows $y(E)$ in Eq.~(\ref{EE}) and $E$ as a function of $E$ with ${\omega _0} = 1.7\Gamma $. (b) denotes $y({\omega _c}) - {\omega _c}$ as a function of ${\omega _0}$. We put $y({\omega _c}) - {\omega _c}<0$ in regimes I and III, and $y({\omega _c}) - {\omega _c} \ge  0$ in regimes II and IV. The number of intersections correspond to the number of bound states when ${y_1}(E)$ and ${y_2}(E)$ (see Eqs.~(\ref{y1}) and (\ref{y2})) change with the eigenvalue $E$, where the parameters chosen are ${\omega _0} = 0.8\Gamma $ for (c), ${\omega _0} = 1.8\Gamma $ for (d), ${\omega _0} = 2.1\Gamma $ for (e), and ${\omega _0} = 3.4\Gamma $ for (f). The other parameters chosen are $\kappa  = 0.59\Gamma $ and ${\omega _c} = 2\Gamma $.}
\label{ss}
\end{figure*}
To clarify this point, we proceed by solving the eigenequation
\begin{equation}
\begin{aligned}
\hat H\left| \varphi \right\rangle  = E\left| {\varphi } \right\rangle
,\label{HE}
\end{aligned}
\end{equation}
where $\hat H$ is given by Eq.~(\ref{Ht}), and the eigenstate
\begin{equation}
\begin{aligned}
\left| \varphi  \right\rangle  &= [{r_A}{\left| e \right\rangle _A}{\left| g \right\rangle _B} + {r_B}{\left| g \right\rangle _A}{\left| e \right\rangle _B}] \otimes {\left| {{0_k}} \right\rangle _E}\\
&\quad+ \sum\nolimits_k {{{\tilde g}_k}} {\left| g \right\rangle _A}{\left| g \right\rangle _B}{\left| {{1_k}} \right\rangle _E}
\label{varphi}
.\end{aligned}
\end{equation}
Substituting Eqs.~(\ref{Ht}) and (\ref{varphi}) into Eq.~(\ref{HE}), we can obtain a set of equations for
\begin{eqnarray}
E{r_A}&=&{\omega _0}{r_A} + \kappa {r_B} + \sum\limits_k {{c_k}{{{\tilde g}_k}}},\label{EH1}\\
E{r_B}&=&{\omega _0}{r_B} + \kappa {r_A},\label{EH2}\\
E{{{\tilde g}_k}}&=&{\omega _k}{{{\tilde g}_k}}+ c_k^ * {r_A}
.\label{EH3}
\end{eqnarray}
Solving Eqs.~(\ref{EH2}) and (\ref{EH3}), we have
\begin{equation}
\begin{aligned}
{r_B} &=& \frac{{\kappa {r_A}}}{{E - {\omega _0}}},\\
{{{\tilde g}_k}} &=& \frac{{c_k^ * {r_A}}}{{E - {\omega _k}}}
.\label{GK}
\end{aligned}
\end{equation}
Substituting Eq.~(\ref{GK}) into Eq.~(\ref{EH1}), we can get the transcendental equation about eigenvalue $E$
\begin{equation}
\begin{aligned}
y(E) \equiv  {\omega _0} + \frac{{{\kappa ^2}}}{{E - {\omega _0}}} - \int_{{\omega _c}}^\infty  {\frac{{{\cal I}(\omega )}}{{\omega  - E}}d\omega } =E
,\label{EE}
\end{aligned}
\end{equation}
where ${{\cal I}(\omega )}$ is given by Eq.~(\ref{Jw}). It is easy to find ${{\omega _0}}$ is a singular point in Eq.~(\ref{EE}). As shown by the blue dash-dotted lines in Fig.~\ref{ss}(a), $y(E)$ is a decreasing function in the interval $\left[ { 0 ,{\omega _0}} \right)$  and also a decreasing function in the interval $\left( {{\omega _0}, {{\omega _c}} } \right]$ (here we set ${\omega _0} = 1.7\Gamma $, ${{\omega _c}}=2\Gamma$), while when ${\omega _0} > {\omega _c}$, there exists only one decreasing function. For bound states to exist in the spectrum of Eq.~(\ref{HE}), Eq.~(\ref{EE}) must have at least one real solution within the energy range. We initially consider a more stringent condition for the formation of bound states. Specifically, if there exists a solution for the third term in Eq.~(\ref{EE}), then we must let $E<\omega_c$.  Therefore, there is one real solution only if the following condition is satisfied: $y({\omega _c}) < {\omega _c}$, i.e.,
\begin{equation}
\begin{aligned}
{\omega _0} + \frac{{{\kappa ^2}}}{{{\omega _c} - {\omega _0}}} - \int_{{\omega _c}}^\infty  {\frac{{{\cal I}(\omega )}}{{\omega  - {\omega _c}}}d\omega }  < {\omega _c}
.\label{wewe}
\end{aligned}
\end{equation}
We plot the line of $y({\omega _c}) - {\omega _c}$ as a function of ${\omega _0}$ in Fig.~\ref{ss}(b), where we have divided into four regimes according to whether the result is greater than or less than 0. It intuitively describes that when $y({\omega _c}) - {\omega _c}<0$, there must be a bound state formed. Conversely, when $y({\omega _c}) - {\omega _c}\ge 0$, no bound state is formed. However, we prove (but not show) that in regime II, the battery energy follows the motion behavior of the red-dashed line in Fig.~(\ref{all})(a), although $y({\omega _c}) - {\omega _c} \ge 0$. This means that there is a bound state in regime II. By exploring the reason, we find $\omega_0<\omega_c$ in regime II. As shown in Fig.~\ref{ss}(a), two decreasing functions exist with ${\omega _0}$ as the boundary. The inequality $y({\omega _c}) - {\omega _c}<0$ indicates that when $E = {\omega _c}$,  $y(E)$ does not intersect $E$ at ${\omega _c}$. However, the left-hand-side decreasing function $y(E)$ bounded by ${\omega _0}$ may intersect $E$. Therefore, Eq.~(\ref{wewe}) is a relatively strict condition for the formation of bound states, and the regime II should also be classified as the regime where a bound state exists. Now we show the number of bound states formed in different regimes and define
\begin{eqnarray}
{y_1}(E) &\equiv& E - \frac{{{\kappa ^2}}}{{E - {\omega _0}}},\label{y1}\\
{y_2}(E) &\equiv& {\omega _0} - \int_{{\omega _c}}^\infty  {\frac{{{\cal I}(\omega )}}{{\omega  - E}}d\omega }
.\label{y2}
\end{eqnarray}
Setting ${y_1}(E) = {y_2}(E)$ with a fixed ${\omega _0}$, we have found different numbers of bound states based on the intersections when ${y_1}(E)$ and ${y_2}(E)$ vary with the eigenvalue as shown in Figs.~\ref{ss}(c)-(f). In regime I of Fig.~\ref{ss}(b), two intersections are observed, corresponding to two bound states. In regimes II and III, there exists a single intersection, which is associated with one bound state. However, in regime IV of Fig.~\ref{ss}(b), no intersection is found, indicating the absence of a bound state formation. Therefore, we consider regimes II and III (e.g., $1.66573\Gamma < {\omega _0} < 3.04138\Gamma$) as an interval for one bound state.

The phase diagram derived from Eq.~(\ref{wewe}) is clear in Fig.~\ref{REDE}(a). Two dashed lines show the first critical equation in Eq.~(\ref{wewe}). The space of bound states about ${\omega _0}$ and $\kappa $ can be clearly found. Hence, we can find that the system-environment bound states are easily controlled by tuning the coupling strength of the interaction between two atoms and the atomic eigenfrequency. Moreover, Fig.~\ref{REDE}(b) more clearly presents that by controlling $\kappa  = 2\Gamma $, both the two bound state regime and the no bound state regime have been transformed into one bound state regime under the original parameters in Fig.~\ref{all}(a), i.e., these battery energy lines exhibit a bound state dynamical behavior. Therefore, it verifies the correctness of our phase diagram. In addition, appropriately increasing ${\omega _0}$ and $\kappa $ within the same regime is beneficial for improving the maximum peak of battery energy or extractable energy as shown in Fig.~\ref{REDE}(b) and Fig.~\ref{REDE}(c).
\end{subsection}
\begin{figure*}[t]
\centering
\includegraphics[width=0.7\textwidth, height=8cm, clip]{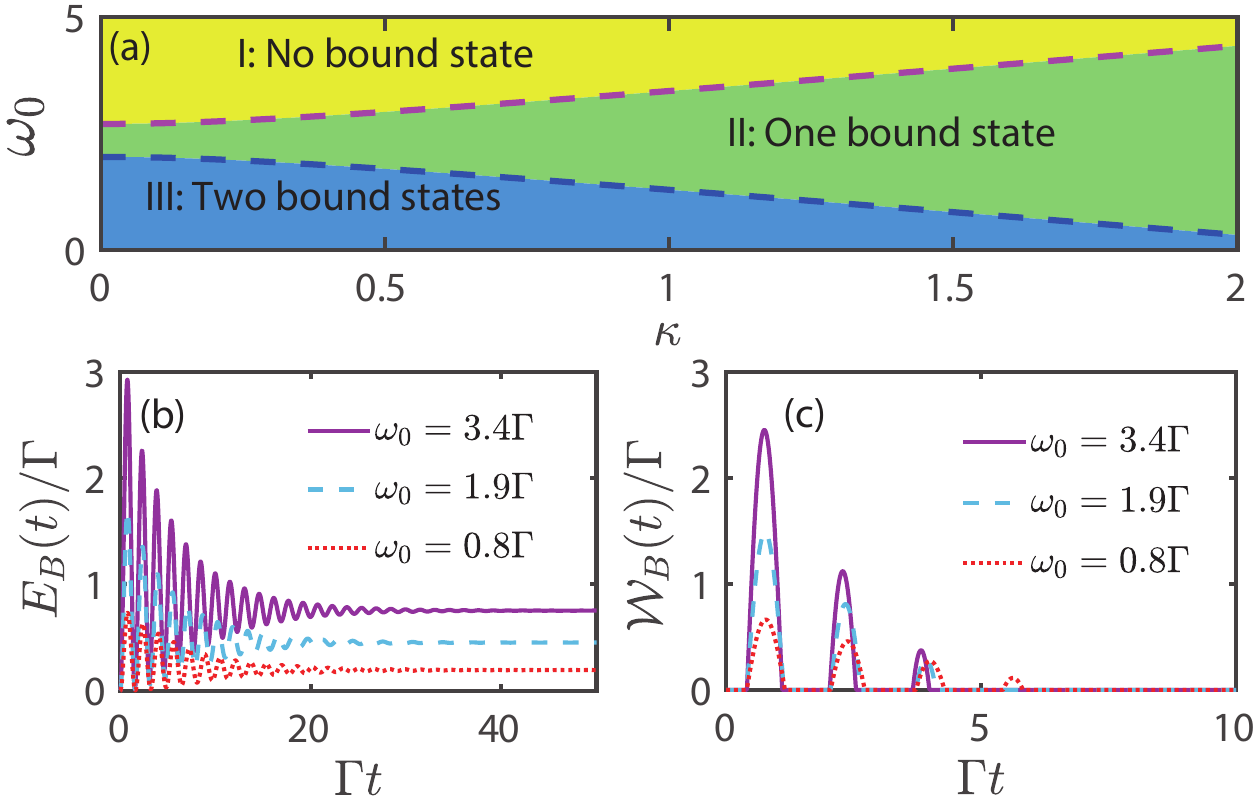}
\caption{(a) Phase diagram of ${\omega _0}$ and $\kappa $ plane. The pink-dashed and blue-dashed lines denote the critical equation in Eq.~(\ref{wewe}), which can be divided into three regimes, i.e., no bound state regime I, one bound state regime II, and two bound states regime III. (b)(c) show that the time evolution of the battery energy ${E_B}(t)$ and the obtained ${{\cal W}_B}(t)$ in Eqs.~(\ref{EB}) and (\ref{WB}) as a function of the dimensionless quantity $\Gamma t$ with different ${\omega _0}$ (in units of $1/\Gamma $). The other parameters chosen are $\kappa  = 2\Gamma $, ${\omega _c} = 2\Gamma $.}
\label{REDE}
\end{figure*}
\\ 
\\
\begin{subsection}{Steady-State Battery Energy and  Ergotropy}
At the end of this section, according to Eq.~(\ref{RAt1}) and Eq.~(\ref{RB}) in Appendix \ref{A}, we focus on the fact that the probability amplitudes ${{\cal R}_B}(t)$ reach a certain state in the long-time regime $(t \to \infty )$. It should be noted that, regarding the sum terms in Eq.~(\ref{RAt1}), the values of the parameters are $M = 0$ and $N = 2$ in the absence of bound states. When there is a single bound state, $M = 1$ and $N = 1$. In the case of two bound states, $M = 2$ and $N = 0$. Thus, we obtain

\begin{widetext}
\begin{eqnarray}
{{\cal R}_{B0}}(t \to \infty ) &=&0,\\
{{\cal R}_{B1}}(t \to \infty )&=&- i\kappa {e^{ - i{\omega _0}t}}\left[\frac{{{e^{(x_1^{(1)} + i{\omega _0})t}} - 1}}{{P'(x_1^{(1)})(x_1^{(1)} + i{\omega _0})}}\right. \left. - \frac{1}{{Q'(x_1^{(2)})(x_1^{(2)} + i{\omega _0})}}\right] + \frac{\kappa }{{2\pi }}{e^{ - i{\omega _0}t}} \nonumber\\
&&\times \int_0^\infty  \left[\frac{1}{{Q( - u - i{\omega _c})}} - \frac{1}{{P( - u - i{\omega _c})}}\right] \frac{1}{{ - u - i{\omega _c} + {\omega _0}}}du ,\label{RB100}
\end{eqnarray}
\begin{align}
{\cal R}_{B2}(t \to \infty ) &= - i\kappa e^{ - i\omega _0 t}\left[\frac{e^{(x_1^{(1)} + i\omega _0)t} - 1}{P'(x_1^{(1)})(x_1^{(1)} + i\omega _0)} + \frac{e^{(x_2^{(1)} + i\omega _0)t} - 1}{P'(x_2^{(1)})(x_2^{(1)} + i\omega _0)}\right] + \frac{\kappa}{2\pi} {e^{ - i\omega _0 t}}\nonumber\\
&\quad\times\int_0^\infty \left[\frac{1}{Q( - u - i\omega _c)} - \frac{1}{P( - u - i\omega _c)}\right] \frac{1}{ - u - i\omega _c + \omega _0}du,\label{RB200}
\end{align}where subscripts 0, 1, and 2 respectively represent the formation of no bound states, one bound state, and two bound states. In our calculation approach adopted herein, only the exponential term of $e$ (persists in oscillating without decay) and the integral term that remains non-zero as $t$ approaches infinity are retained. The other integral terms are not retained due to the Lebesgue-Riemann lemma \cite{Bochner1949}. Eventually, we substitute Eqs.~(\ref{RB100}) and (\ref{RB200}) into Eq.~(\ref{EB}) to obtain the steady-state battery energy
\begin{eqnarray}
{E_{B1}}(t \to \infty ) &=& {\omega _0}{\left| {{R_{B1}}(t \to \infty )} \right|^2},\label{EB1}\\
{E_{B2}}(t \to \infty ) &=& {\omega _0}{\left| {{R_{B2}}(t \to \infty )} \right|^2},\label{EB2}
\end{eqnarray}
and the corresponding steady-state ergotropy
\begin{eqnarray}
{{\cal W}_{B1}}(t \to \infty ) &=& {\omega _0}[2{\left| {{R_{B1}}(t \to \infty )} \right|^2} - 1] \Theta ({\left| {{R_{B1}}(t \to \infty )} \right|^2} - \frac{1}{2})\label{cWB1},\\
{{\cal W}_{B2}}(t \to \infty ) &=& {\omega _0}[2{\left| {{R_{B2}}(t \to \infty )} \right|^2} - 1] \Theta ({\left| {{R_{B2}}(t \to \infty )} \right|^2} - \frac{1}{2})
.\label{WB2}
\end{eqnarray}
\end{widetext}It is worth noting that the steady-state battery energy in the presence of two bound states still oscillates steadily over time, as shown in Fig.~\ref{EBt00}(a). It matches well with the short-term battery energy, indicating that the battery in the two bound states can achieve long-term periodic storage. It can also be extracted periodically as shown in Fig.~\ref{EBt00}(b).
For the case where a bound state exists, we find that the steady-state battery energy is a constant value. Then, we plot ${E_B}(t \to \infty )$ with ${\omega _0}$ in Fig.~\ref{EBt00}(c) according to Eqs.~(\ref{RB100}) and (\ref{EB1}). It can be observed that when the coupling strength $\kappa $ is fixed, the atomic eigenfrequency ${\omega _0}$ controls the formation of the bound state. When a bound state exists, battery energy is always present. However, we have not yet found a steady-state battery that can be extracted in the presence of a bound state, even with the increment of ${\omega _0}$ and $\kappa $. Therefore, the two bound state regimes in the parameter space in Fig.~\ref{REDE}(a) are the best choices we recommend for long-term storage and extraction.
\begin{figure}[t]
\centerline{
\includegraphics[width=8.0cm, height=7.0cm, clip]{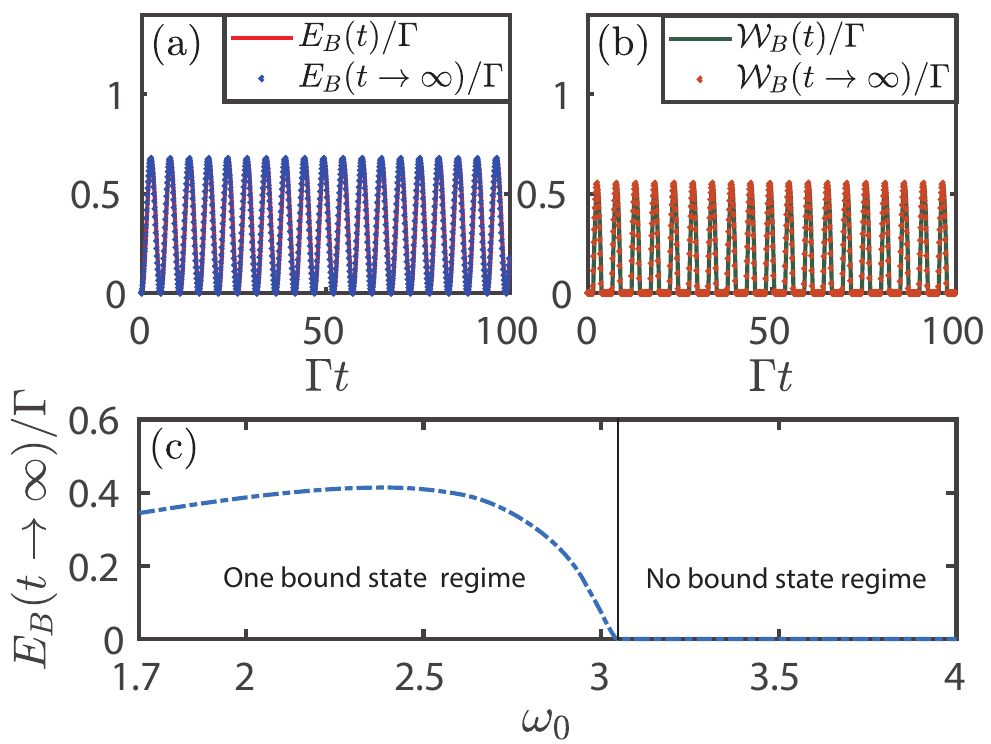}}
\caption{(a) and (b) present the variation of battery energy ${E_B}(t)$ in Eq.~(\ref{EB}), steady-state battery energy ${E_{B2}}(t \to \infty )$ in Eq.~(\ref{EB2}), ergotropy ${\cal W}_B(t)$ in Eq.~(\ref{WB}), and steady-state ergotropy ${\cal W}_{B2}(t \to \infty )$ in Eq.~(\ref{WB2}) over time in the presence of two bound states. (c) shows the steady-state battery energy given by Eq.~(\ref{EB1}) as functions of ${\omega _0}$ in a bound state regime and no bound state regime. The other parameters chosen are ${\omega _c} = 2\Gamma $ and $\kappa=0.59 \Gamma $.}\label{EBt00}
\end{figure}

\end{subsection}

\section{Self-Discharging Process of QB}\label{Self-Discharging Process of Quantum Battery}

In classical battery, there is a known phenomenon as battery self-discharge, which is not desired as it affects storage performance. The self-discharging process is characterized by a loss of charge in the battery, even when the battery is not connected to any consumption device. In the quantum field, the self-discharging mechanism has been introduced due to the inevitable coupling of the QB with some external environment. Now we are studying the self-discharge of energy transferred from the charger to QB after charging is completed in a non-Markovian environment. Hence, the total Hamiltonian is given by
\begin{equation}
\begin{aligned}
{{\hat H}'} = \hat H_0' + \hat H_I',
\label{HHH}
\end{aligned}
\end{equation}
with
\begin{equation}
\begin{aligned}
{{\hat H'}_0} &= {\omega _0}\hat \sigma _B^ + \hat \sigma _B^ -  + \sum\limits_k {{\omega _k}} \hat a_k^\dag {{\hat a}_k},\\
{{\hat H'}_I} &= \sum\limits_k {({g_k}} \hat \sigma _B^ + {{\hat a}_k} + g_k^*\hat \sigma _B^ - \hat a_k^\dag )
,\label{HHH1}
\end{aligned}
\end{equation}
where ${{g_k}}$ is the coupling strength of the interaction between the battery and the $k$th field mode of the environment. We assume that the battery is removed from the charger immediately after it is fully charged, and the evolved state in the total system dynamics can be written as
\begin{equation}
\begin{aligned}
\left| {\Phi (t)} \right\rangle  = {C_1}(t)\left| {e,0} \right\rangle  + \sum\limits_k {{C_{2,k}}} (t)\left| {g,{1_k}} \right\rangle
,\label{phi11}
\end{aligned}
\end{equation}
with normalized condition $|{C_1}(t){|^2} + |\sum\nolimits_k {{C_{2,k}}(t)} {|^2} = 1$.
\begin{figure}
\centerline{
\includegraphics[width=8.5cm, height=3.9cm, clip]{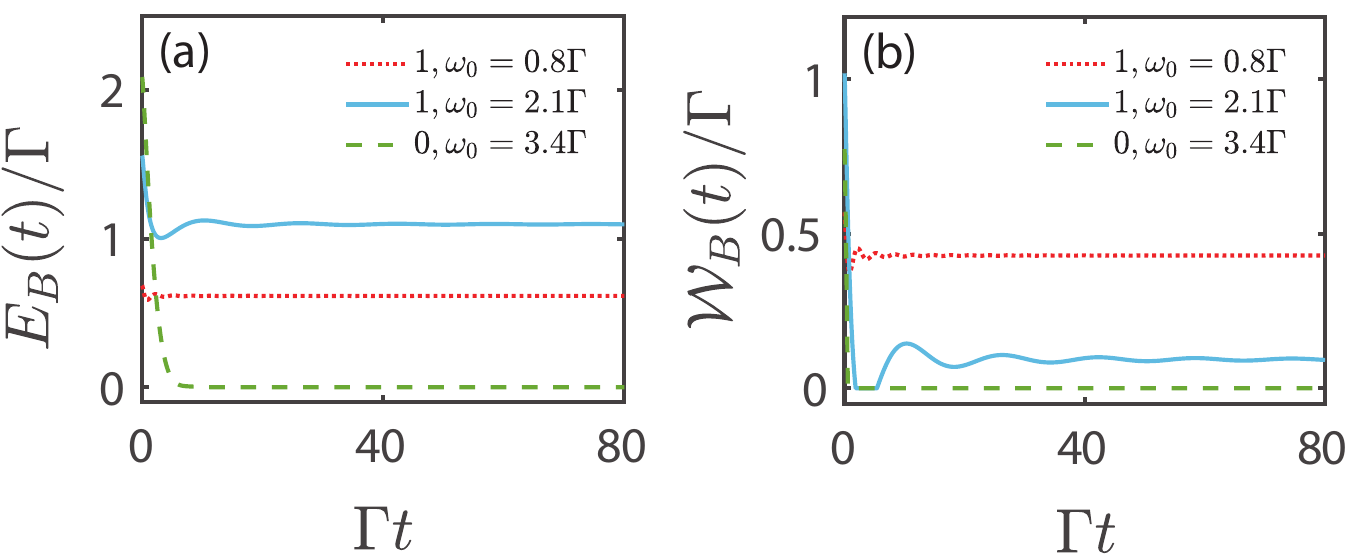}}
\caption{(a) and (b) show that the influences of the bound state on the battery energy ${E_B}(t)$ and the corresponding ergotropy given by Eqs.~(\ref{EBt1}) and (\ref{WBtt}) under self-discharging as a function of the dimensionless quantity $\Gamma t$ with different ${\omega _0}$ (in units of $1/\Gamma $). The blue-solid  and red-dotted lines correspond to scenarios with one bound state. The green-dashed lines represent a case with no bound states. The other parameter chosen is ${\omega _c} = 2\Gamma $.}\label{dEB}
\end{figure}
Using the Schr\"{o}dinger equation and the calculation method of the Appendix \ref{A}, we can obtain
\begin{equation}
\begin{aligned}
{{\dot C}_1}(t) =  - i{\omega _0}{C_1}(t) - \int_0^t {f(t - \tau )} {C_1}(\tau )d\tau
,\label{dCa}
\end{aligned}
\end{equation}
where $f(t - \tau ) = \sum\nolimits_k {{{\left| {{g_k}} \right|}^2}} {e^{ - i{\omega _k}(t - \tau )}} \equiv \int {I(\omega )} {e^{ - i{\omega _k}(t - \tau )}}$ with $I(\omega )\equiv {\cal I}(\omega )$ given by Eq.~(\ref{Jw}). The probability amplitude ${C_1}(t)$ for the battery is
\begin{equation}
\begin{aligned}
{C_1}(t) &= {C_1}(0) \Bigg\{  \sum\limits_M \frac{e^{X_M^{(1)}t}}{G'(X_M^{(1)})} + \sum\limits_N \frac{e^{X_N^{(2)}t}}{L'(X_N^{(2)})} \\
&\quad+ \frac{1}{2\pi i} \int_0^\infty dy \, h_2(-y - i\omega_c) e^{-yt - i\omega_c t} \Bigg\},
\label{cat}
\end{aligned}
\end{equation}
where $G(s)$ and $L(s)$ are given by Eqs.~(\ref{Gs}) and (\ref{Ls}). Here, the quantities in Eq.~(\ref{cat}) are similar to those in the previous equation (\ref{RAt1}) with new letters for the variables. Derivation of Eq.~(\ref{cat}) can be found in Appendix \ref{E}. According to Eq.~(\ref{phi11}), the reduced density operator of the QB at the time $t$ can be obtained as
\begin{equation}
\begin{aligned}
{\rho _B}(t) = \left( {\begin{array}{*{20}{l}}
{{{\left| {{C_1}(t)} \right|}^2}\qquad 0}\\
{\quad 0\qquad 1 - {{\left| {{C_1}(t)} \right|}^2}}
\end{array}} \right)
,\label{rhoB1}
\end{aligned}
\end{equation}
which leads to the battery energy
\begin{equation}
\begin{aligned}
{E_B}(t) = {\omega _0}{\left| {{C_1}(t)} \right|^2}
,\label{EBt1}
\end{aligned}
\end{equation}
and corresponding ergotropy
\begin{equation}
\begin{aligned}
{{\cal W}_B}(t) =  {\omega _0}[2{\left| {{C_1}(t )} \right|^2} - 1]\Theta ({\left| {{C_1}(t )} \right|^2} - \frac{1}{2})
.\label{WBtt}
\end{aligned}
\end{equation}
It is worth noting that if we drive the charger and battery system without structured environment, one finds the required shortest time ${\tau _s}$ to fully transfer the energy from charger to QB and we can obtain ${E_B}({\tau _s}) = {E_{B\max }} = {\omega _0}$, where the maximum energy can be stored in the QB. When we consider the environmental effects on the system, the maximum energy can be changed (e.g., the first peak of the green-solid line in Fig.~\ref{all}(a), where the maximum energy ${E_{B\max }} < {\omega _0}$ ). We assume that the battery is removed when it reaches the maximum energy in the shortest time, i.e., ${\left| {{{\cal R}_B}({\tau _s})} \right|^2} = {\left| {{C_1}(0)} \right|^2}$, and we can get the initial value ${C_1}(0) = 0.915$ for ${\omega _0} = 0.8\Gamma $, ${C_1}(0) = 0.812$ for ${\omega _0} = 2.1\Gamma $, and ${C_1}(0) = 0.714$ for ${\omega _0} = 3.4\Gamma $. Therefore, the time evolution of battery energy can be obtained according to Eq.~(\ref{EBt1}) as shown in Fig.~\ref{dEB} (a). By comparing lines of battery energy, the formation of a bound state may initially lead to the loss of battery energy. Subsequently, the environmental exciton enters the battery and finally reaches a state without self-discharge. It should be noted that there exists only one bound state in the battery-environment system given by Eq.~(\ref{HHH}) compared to Eqs.~(\ref{DRAt}) and (\ref{dCa}) or Eqs.~(\ref{EE}) and (\ref{yEP}) (we prove that the two functions $y(E')$ and ${E'}$ corresponding to the transcendental equation have only one intersection point, indicating that the whole system has formed one bound state revealed in Appendix \ref{E}). The green-dashed line corresponds to no bound state, which shows that the battery will gradually discharge until it becomes empty, i.e., the battery energy is zero.

 Meanwhile, we discover that energy extraction is feasible even when the battery forms only one bound state in this scenario. This is because the value of the battery energy at the end of the charging process is selected as the relatively high maximum peak energy. Such a value enables the steady-state energy of the ultimately forming single bound state to reach an extractable condition as shown in Fig.~\ref{dEB}(b). We also find the parameters that form a two-bound-state regime during charging are optimal choice for extraction situation after long-term storage (see the red-dotted line in Fig.~\ref{dEB}(b)). Therefore, during the self-discharge process, even if only a single bound state is formed, it can still significantly enhance the energy storage and extraction performance of QB. Overall, the formation of bound states reveals their universal role in suppressing decoherence.

\section{Conclusions and Discussions}\label{Conclusions and Discussions}

In conclusion, we have established that the formation of system-environment bound states serves as a  highly effective strategy to suppress decoherence in quantum batteries. By exactly solving the non-Markovian dynamics of a charger-battery system coupled to a 3D photonic crystal, we have demonstrated a direct mapping between the spectral properties of the environment—specifically the number of bound states—and the macroscopic performance of the battery. Our most significant finding is the identification of a two-bound-state regime, where the battery energy does not decay but instead maintains a persistent periodic oscillation. This represents a fundamental paradigm shift: instead of merely delaying energy loss, we show that permanent energy storage and cyclic extraction are theoretically achievable without external interference. Furthermore, our derived phase diagram provides experimentalists with a clear ``control map" to navigate between rapid charging (short-term) and durable storage (long-term) simply by adjusting the atomic eigenfrequency and coupling strength.

Looking forward, our results pave the way for the realization of robust, long-lifetime quantum energy devices. The exact solvability of our model also opens avenues to extend these insights to non-rotating-wave approximations \cite{Shen042121,Shen023856,Shen013826,Shen043714} and other complex spectral densities (e.g., Ohmic spectra) \cite{Wei050401,Yang022122,Groblacher7606}. We believe this work not only deepens the fundamental understanding of open quantum systems but also accelerates the practical deployment of quantum batteries in realistic photonic and superconducting platforms.

\section*{ACKNOWLEDGMENTS}
This work was supported by Science and Technology Development Plan Project of Jilin Province (Grant No.~20250102007JC), National Natural Science Foundation of China (NSFC) under Grants No.~12274064 and No. 12374333, and the Hakubi Projects of RIKEN.

\section*{DATA AVAILABILITY}
The data that support the findings of this article are not
publicly available. The data are available from the authors
upon reasonable request.

\section{Appendix} \par

\subsection{\label{A} The Derivation of Integro-Differential Equation for Amplitude ${{\cal R}_A}(t)$.}
Through the Schr\"{o}dinger equation $i\frac{d}{{dt}}\left| {\psi (t)} \right\rangle  = \hat H\left| {\psi (t)} \right\rangle $ and the initial condition $\left| {\psi (0)} \right\rangle  = {\left| e \right\rangle _A}{\left| g \right\rangle _B}{\left| {{0_k}} \right\rangle _E}$ with ${{\cal R}_A}(0) = 1$, ${{\cal R}_B}(0)=0$, and ${{{\cal G}_k}(0)}=0$, the probability amplitudes are given by the differential equation
\begin{align}
\frac{d}{{dt}}{{\cal R}_A}(t) &=- i[{\omega _0}{{\cal R}_A}(t) + \kappa {{\cal R}_B}(t) + \sum\limits_k {{c_k}{{\cal G}_k}(t)} ],\label{dRA}\\
\frac{d}{{dt}}{{\cal R}_B}(t) &=- i[{\omega _0}{{\cal R}_B}(t) + \kappa {{\cal R}_A}(t)],\label{dRB}\\
\frac{d}{{dt}}{{\cal G}_k}(t) &=- i[c_k^ * {{\cal R}_A}(t) + {\omega _k}{{\cal G}_k}(t)]
.\label{dG}
\end{align}
We solve Eq.~(\ref{dRB}) with the initial conditions ${{\cal R}_B}(0) = 0$ and get
\begin{equation}
\begin{aligned}
{{\cal R}_B}(t) =  - i\kappa \int_0^t {{{\cal R}_A}(\tau )} {e^{ - i{\omega _0}(t - \tau )}}d\tau
.\label{RB}
\end{aligned}
\end{equation}
By using the differential equation (\ref{dG}) with the initial conditions ${{\cal G}_k}(0) = 0$, the probability amplitude ${{\cal G}_k}(t)$ can be written as
\begin{equation}
\begin{aligned}
{{\cal G}_k}(t) =  - i\int_0^t {c_k^ * } {{\cal R}_A}(\tau ){e^{ - i{\omega _k}(t - \tau )}}d\tau
.\label{Gkt}
\end{aligned}
\end{equation}
Substituting Eq.~(\ref{RB}) and Eq.~(\ref{Gkt}) into Eq.~(\ref{dRA}), we can obtain Eq.~(\ref{DRAt}).

\subsection{\label{D} The Derivation of Exact Solution for Amplitude ${{\cal R}_A}(t)$ and Ergotropy}
To grasp qualitatively the physics behind the threshold in the non-Markovian effect, we solve Eq.~(\ref{DRAt}) by Laplace transformation that yields
\begin{equation}
\begin{aligned}
{{\cal R}_A}(s) = \frac{{{{\cal R}_A}(0)}}{{s + i{\omega _0} + {\cal M}(s) + {\cal F}(s)}}
,\label{RAs}
\end{aligned}
\end{equation}
where
\begin{eqnarray}
{\cal M}(s) &=& \frac{{{\kappa ^2}}}{{s + i{\omega _0}}},\label{ms}\\
{\cal F}(s) &=& \int {\frac{{{\cal I}(\omega )}}{{s + i\omega }}} d\omega
.\label{Fs}
\end{eqnarray}
By substituting Eq.~(\ref{Jw}) into Eq.~(\ref{Fs}), we can get correlation function in the frequency domain
\begin{equation}
\begin{aligned}
{\cal F}(s) = \frac{{ - i{\Gamma ^{3/2}}}}{{\sqrt {{\omega _c}}  + \sqrt { - is + {\omega _c}} }}
,\label{Fs1}
\end{aligned}
\end{equation}
where the phase angle of $s$ is defined by $ - \pi  < \arg (s) < \pi $; the phase angle of ${\sqrt { - is + {\omega _c}} }$ in ${\cal F}(s)$ is defined by $ - \frac{\pi }{2} < \arg \sqrt { - is + {\omega _c}}  < \frac{\pi }{2}$. The excited-state amplitudes ${{\cal R}_A}(t)$ of atom $A$ can then be acquired through an inverse Laplace transformation
\begin{equation}
\begin{aligned}
{{\cal R}_A}(t) = \frac{1}{{2\pi i}}\int_{\sigma  - i\infty }^{\sigma  + i\infty } {{{\cal R}_A}(s)} {e^{st}}ds
,\label{RAt11}
\end{aligned}
\end{equation}
where ${{\cal R}_A}(s) = \frac{{{R_A}(0)}}{{s + i{\omega _0} + {\cal M}(s) + {\cal F}(s)}}$. With the integration contours as shown in black-solid line of
Fig.~(\ref{s}), we have
\begin{widetext}
\begin{equation}
\begin{aligned}
{{\cal R}_A}(t) =& \sum\limits_M {\frac{{{{\cal R}_A}(0){e^{x_M^{(1)}t}}}}{{P'(x_M^{(1)})}} - \frac{1}{{2\pi i}}} \Bigg\{ \int_{ - i{\omega _c} - \infty }^{ - i{\omega _c} + 0} { + \int_{ - i{\omega _c} + 0}^{ - i\infty  + 0} {ds} }  \frac{{{{\cal R}_A}(0)}}{{s + i{\omega _0} + {\cal M}(s) + {\cal F}(s)}}{e^{st}}\Bigg\}
,\label{RAt2}
\end{aligned}
\end{equation}
\end{widetext}
where the function
\begin{equation}
\begin{aligned}
P(s) = s + i{\omega _0} + {\cal M}(s)+ {\cal F}(s)
,\label{Ps}
\end{aligned}
\end{equation}
and ${x_M^{(1)}}$ are the roots of equation $P(s) = 0$ in the regime [${\mathop{\rm Re}\nolimits} (s) > 0$ or ${\mathop{\rm Im}\nolimits} (s) >  - {\omega _c}$]. ${\mathop{\rm Re}\nolimits} (s) = \sigma $ lies to the right of all the singularities ${x_M^{(1)}}$. The calculation of the last term employs the integral contours indicated by the black-dashed line as depicted in Fig. (\ref{s})
\begin{widetext}
\begin{equation}
\frac{1}{{2\pi i}}\int_{ - i{\omega _c} + 0}^{ - i\infty  + 0} {ds} \frac{{{{\cal R}_A}(0)}}{{s + i{\omega _0} + {\cal M}(s) + {\cal F}(s)}}{e^{st}}= - \sum\limits_{\cal M} {\frac{{{{\cal R}_A}(0){e^{x_N^{(2)}t}}}}{{Q'(x_N^{(2)})}}}  - \frac{1}{{2\pi i}}\Bigg\{ \int_{ - i{\omega _c} - \infty }^{ - i{\omega _c} + 0} {ds}  \frac{{{{\cal R}_A}(0)}}{{s + i{\omega _0} + {\cal M}(s) + {\cal F}(s)}}{e^{st}}\Bigg\}
,
\end{equation}
\end{widetext}
where the function
 \begin{equation}
\begin{aligned}
 Q(s) = s + i{\omega _0} +{\cal M}(s)+ {\cal F}_1(s),
 \label{Qs}
\end{aligned}
\end{equation}
 with ${\cal F}_1(s) =  - i{\Gamma ^{3/2}}/(\sqrt {{\omega _c}}  - i \sqrt {  is - {\omega _c}} )$. ${x_N^{(2)}}$ are the roots of $Q(s) = 0$ in regime [${\mathop{\rm Re}\nolimits} (s) < 0$ and ${\mathop{\rm Im}\nolimits} (s) <  - {\omega _c}$].  From Eqs. (\ref{RAt11}), (\ref{Ps}), and (\ref{Qs}), we can obtain the amplitude (\ref{RAt1}) by setting $s =  - u - i{\omega _c}$. Regarding the specific calculation of ergotropy, we define ${\sigma _\rho }$, called the passive state, with zero extractable energy by cyclic unitary operations. Then, considering a reference Hamiltonian ${\hat H}$ and a given state ${\rho}$, we can write both of them in their respective eigenbasis (increasing for ${\hat H}$, decreasing for ${\rho}$) \cite{Alicki87042123,Kamin22083007}
\begin{eqnarray}
\hat H = \sum\limits_i {{E_i}\left| {{E_i}} \right\rangle } \left\langle {{E_i}} \right|, \quad {E_{i + 1}} \ge {E_i} \quad \forall i,\label{Hi}\\
{\rho} = \sum\limits_i {{w_i}\left| {{w_i}} \right\rangle } \left\langle {{w_i}} \right|,\quad {w_{i + 1}} \le {w_i} \quad  \forall i
.\label{wi}
\end{eqnarray}
The state $\rho $ is passive with respect to ${\hat H}$ if and only if (i) $\rho $ and ${\hat H}$ are diagonal in the same basis being valid with the commutation relation ${\rm{[}}\rho ,\hat H{\rm{] = 0}}$, and (ii) $\rho $ contains no population inversion, i.e., ${E_i} < {E_{i + 1}} \Rightarrow {w_i} \ge {w_{i + 1}}$. Thus the passive state takes the following form
\begin{equation}
\begin{aligned}
{\sigma _\rho }  = \sum\limits_i {{w_i}\left| {{E_i}} \right\rangle } \left\langle {{E_i}} \right|
.\label{passive}
\end{aligned}
\end{equation}
According to Eqs.~(\ref{WB1}) and (\ref{passive}), the ergotropy can be obtained as
\begin{equation}
\begin{aligned}
{{\cal W}_B}(t) = {\rm{Tr}}[{\rho _B}(t)\hat H_0^B]{\rm{ - Tr}}[{\sigma _{{\rho _B}}(t)}\hat H_0^B].
\label{WBt}
\end{aligned}
\end{equation}
Specifically, we have quantities of battery (atom $B$) expanded in terms of eigenbases as follows
\begin{equation}
\begin{aligned}
\hat H_0^B =& 0\left| g \right\rangle \left\langle g \right| + {\omega _0}\left| e \right\rangle \left\langle e \right|,\\
{\rho _B}(t) =& {\left| {{{\cal R}_B}(t)} \right|^2}\left| e \right\rangle \left\langle e \right| + (1 - {\left| {{{\cal R}_B}(t)} \right|^2})\left| g \right\rangle \left\langle g \right|,\\
{\sigma _{{\rho _B}}(t)} =& {\left| {{{\cal R}_B}(t)} \right|^2}\left| g \right\rangle \left\langle g \right| + (1 - {\left| {{{\cal R}_B}(t)} \right|^2})\left| e \right\rangle \left\langle e \right|,
\end{aligned}
\end{equation}
where it is required ${| {{{\cal R}_B}(t)} |^2} \ge 1 - {| {{{\cal R}_B}(t)} |^2} \Rightarrow {| {{{\cal R}_B}(t)} |^2} \ge 0.5$. Substituting the above equations into Eqs. (\ref{WBt}), the ergotropy in Eq. (\ref{WB}) can be got.

\subsection{\label{E} The Derivation of Eq.~(\ref{cat}) and Bound State Situation under Self-Discharge Case}
Solving Eq.~(\ref{dCa}) by Laplace transformation, we have
\begin{equation}
\begin{aligned}
{C_1}(s) = \frac{{{C_1}(0)}}{{s + i{\omega _0} + F(s)}} \equiv \frac{{{C_1}(0)}}{{G(s)}},\label{calg}
\end{aligned}
\end{equation}
where the function
\begin{equation}
\begin{aligned}
G(s) = s + i{\omega _0} + {\cal F}(s),
\label{Gs}
\end{aligned}
\end{equation}
with ${\cal F}(s) =  - i{\Gamma ^{3/2}}/(\sqrt {{\omega _c}}  + \sqrt { - is + {\omega _c}} )$. Defining
\begin{equation}
\begin{aligned}
L(s) = s + i{\omega _0} + {{\cal F}_1}(s),
\label{Ls}
\end{aligned}
\end{equation}
with ${{\cal F}_1}(s) =  - i{\Gamma ^{3/2}}/(\sqrt {{\omega _c}}  - i\sqrt {is - {\omega _c}} )$, and applying the calculation method of Appendix \ref{D}, we can obtain Eq.~(\ref{cat}).

\begin{figure}[t]
\centerline{
\includegraphics[width=4.5cm, height=4.5cm, clip]{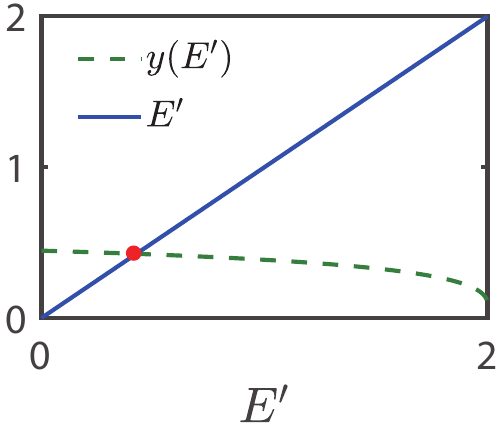}}
\caption{The figure shows the intersection of functions $y(E')$ and $E'$ at a point, where $y(E')$ is a decreasing function. The other parameters chosen are ${\omega _0} = 0.8\Gamma $ and ${\omega _c} = 2\Gamma $.} \label{disyE}
\end{figure}
Now we study the energy spectrum of the self-discharge case. The eigenequation is as follows
\begin{equation}
\begin{aligned}
\hat H'| {\tilde \Phi } \rangle   = E'| {\tilde \Phi } \rangle
,\label{HP}
\end{aligned}
\end{equation}
where $\hat H'$ is given by Eq.~(\ref{HHH}) and the corresponding eigenstate is
\begin{equation}
\begin{aligned}
| {\tilde \Phi } \rangle  = {{\tilde c}_1}| {e,0} \rangle  + \sum\limits_k {{{\tilde c}_{2,k}}}| {g,{1_k}} \rangle
.
\end{aligned}
\end{equation}
Through some calculations, we obtain the transcendental equation under the self-discharge case
\begin{equation}
\begin{aligned}
y(E') \equiv {\omega _0} - \int_{{\omega _c}}^\infty  {\frac{{{\cal I}(\omega )}}{{\omega  - E'}}} d\omega = E'
,\label{yEP}
\end{aligned}
\end{equation}
where $ \cal{I}(\omega )$ can be found in Eq.~(\ref{Jw}). In contrast to Eq.~(\ref{EE}), $\omega_0$ no longer represents a singular point. Therefore, the function $y(E')$ is strictly a decreasing function. Consequently, there exists only one intersection point between functions $y(E')$ and $E'$ as illustrated in Fig.~\ref{disyE}. We take the parameters that fall in a two-bound-state regime when charging, i.e., ${\omega _0} = 0.8\Gamma$ and ${\omega _c} = 2\Gamma$ corresponding to Fig.~\ref{ss} (c). It can be observed that the two functions of the transcendental equation only have one intersection point in the self-discharge case. This indicates that the entire system only forms one bound state.


\end{document}